\documentclass[12pt]{krantz}
\usepackage{makeidx}
\usepackage{amsmath, amssymb,amsbsy}
\usepackage{lipsum}
\usepackage{xspace}
\usepackage{algorithmic, algorithm}
\usepackage{latexsym}
\usepackage{enumerate}
\usepackage[T1]{fontenc}   %%%%ST changed
\usepackage{bm}
\usepackage{rotating} % For landscape figures and tables
\usepackage{booktabs} % For table
\usepackage{multicol,dcolumn}
\usepackage{threeparttable} % This allows notes in tables
 \usepackage{natbib}   % This is a great aid with bibliographies
\usepackage{url}      % This package helps to typeset urls
\usepackage{caption}
\usepackage{subcaption} % This allows multiple captioned plots within a figure 

\newtheorem{definition}{Definition}
\newtheorem{example}{Example}

\def\bhg{\hangindent=10pt\hangafter=1}

\def\bhg1{\hangindent=12pt\hangafter=1}

\newcommand{\reals}{\mbox{\rm I\kern-.20em R}}
\newcommand{\bX}{{\mathbf X}}

\newcommand{\bx}{{\mathbf x}}

\newcommand{\bY}{{\mathbf Y}}
\newcommand{\by}{{\mathbf y}}

\newcommand{\bu}{{\mathbf u}}
\newcommand{\bU}{{\mathbf U}}

\newcommand{\widesim}[2][1.5]{
  \mathrel{\overset{#2}{\scalebox{#1}[1]{$\sim$}}}
}
\begin{document}

\frontmatter
\title{Bridging Bayesian, frequentist and fiducial %(BFF)
inferences using confidence distributions}
\author{Suzanne Thornton %Department of Mathematics and Statistics, Swarthmore College  %Swarthmore, PA 19081, USA 
and Min-ge Xie} %  Department of Statistics, Rutgers University}% \\ New Brunswick, NJ 08854, USA}
\maketitle

\footnotetext[12]{ 
Email addresses: sthornt1@swarthmore.edu and mxie@stat.rutgers.edu.  The research is supported in part by NSF research grants: DMS1737857, DMS1812048, DMS2015373 and DMS2027855}

\setcounter{page}{7} %%% Set page number to 7

\tableofcontents

\mainmatter
\chapter{Bridging Bayesian, frequentist and fiducial inferences using confidence distributions}

\section{Introduction}\label{sec:intro}

Bayesian, frequentist, and fiducial (BFF) inferences are much more congruous than have been perceived historically in the scientific community (e.g., \citet{Reid2015,  Kass2011,Efron1998}). 
Most practitioners are probably more familiar with the two dominant statistical inferential paradigms, Bayesian inference and frequentist inference. 
The third, lesser known fiducial inference paradigm was pioneered by R.A. Fisher in an attempt to define  an inversion procedure for inference as an alternative to Bayes' theorem.  
Although each paradigm has its own strengths and limitations subject to 
their different philosophical underpinnings, this article intends to bridge these different inferential methodologies %through the lenses of 
by calling upon {\it confidence distribution} theory and Monte-Carlo simulation procedures, thereby increasing the range of possible techniques available to both statistical theorists and practitioners across all fields.

One motivation behind the development of confidence distributions is to address a simple question: ``Can frequentist inference rely on a distribution function, or a {\it distribution estimator}, in a manner similar to a Bayesian posterior?'' (e.g., \citealp{Xie2013a}). The notion of a confidence distribution answers this question in the affirmative.
%; in fact,
% m
Many statisticians have considered confidence distributions to be a frequentist analogue to a Bayesian posterior \citep[chap. 2]{Schweder2003b}. 
Although by definition, a confidence distribution is defined within a frequentist framework, this special type of estimator can potentially unite Bayesian, frequentist, and fiducial inferential methods. In the subsequent sections we provide a brief review of the concept of a confidence distribution and illustrate 
%develop an appreciation for 
the flexibility of inferential framework this provides. 
We further %also
highlight two deeply rooted connections 
% connections 
involving the estimation and uncertainty quantifications among confidence distributions, bootstrap distributions, fiducial distributions, and Bayesian posterior distributions. 
Here, we use the term ``uncertainty'' to specifically refer to the variability associated with the random sample under an assumed model, called ``empirical variability'' in \citet{Reid2015}.
We present a unified platform from which one may compare and combine inference across different paradigms and derive 
new simulation-based inference approaches. 
We conclude by presenting a novel assertion that, across all inferential paradigms, a parameter has both a ``fixed'' and a ``random'' version.

\section{Confidence Distribution (CD): A Distribution Estimator}\label{sec:distributionest} 
\subsection{Definitions}\label{sec:definition}
Traditionally, statistical analysis uses a single point or perhaps an interval to estimate an unknown target parameter. However, another valid option is to consider %using a 
a sample-dependent function, specifically, a sample-dependent function on the parameter space, to estimate the parameter of interest. This is precisely the purpose of a confidence distribution. A confidence distribution is a sample-dependent, distribution (or density) function whose main purpose is 
estimation, including an assessment of uncertainty, %in the estimation of 
for an unknown target parameter, similar to 
any other statistical estimator. 
A confidence distribution can be defined using a simple, prescriptive definition. In the definition below and throughout the remainder of this chapter we use $\mathcal{Y}$ and $\Theta$ to denote the sample space and parameter space, respectively. We also use the conventional notation to clarify the difference between random variables, $\bY$, and their observations,~$\by$.

\begin{definition} A sample-dependent function on the parameter space, i.e. a function on $\mathcal{Y} \times \Theta$, $H_{n}(\cdot) = H_{n}({\bY}, \cdot)$, is called a \emph{confidence distribution (CD)} for $\theta \in \Theta$ if \\ 
{[R1]} For each given sample $\by\in \mathcal{Y}$, the function $H_{n}(\cdot) = H_{n}({\by}, \cdot)$ is a distribution function on the parameter space $\Theta$; and \\
{\it [R2]} The function can provide confidence intervals (or regions) for the parameter $\theta$ at all confidence levels.

If [R2] only holds asymptotically (or approximately), then $H_{n}(\cdot)$ is called an \emph{asymptotic (or approximate) CD (aCD)}.    
\end{definition}\label{def:CD}

In plain language, a CD is a function of both the parameter and the random sample with two requirements: {\it [R1]} that the estimator is a sample-dependent (distribution) function on the parameter space and {\it [R2]} that a particular performance property holds. This type of definition is analogous to the way consistent (or unbiased) estimators are defined in point estimation. Specifically, consistent (or unbiased) point estimators are determined as such if first ({\it [R1]}), the estimator is a sample-dependent point in the parameter space and second ({\it [R2]}), the estimator satisfies a particular performance-based criteria. For a consistent estimator, the latter ({\it [R2]}) requirement is that the estimator approaches the true parameter value as the sample size increases; for unbiased estimators this requirement is instead that the expectation of the estimator is equal to the true parameter value. For the distributional estimator CD, the requirement {\it [R2]} instead states that this estimator must provide valid confidence intervals (or regions) for any significance level.

When the parameter $\theta$ is scalar, requirement {\it [R2]} in the CD definition can be restated %expressed 
as the requirement that, at the true parameter value $\theta=\theta_0$, $H_{n}(\theta_0) \equiv H_{n}(\bY, \theta_0)$ as a function of the random data follows a standard uniform distribution (e.g., \citet{Xie2013a}; \citet[chap. 3]{Schweder2016}). The classical definition of CD in \citet{Cox1958} and \citet{Efron1993} is obtained by inverting the upper endpoints of a set of nested one-sided confidence intervals of all levels. This classical definition is covered by the descriptive definition of a CD in Definition 1 (e.g., \citet[\textsection 2.1 \& 2.2]{Xie2013a}).  
When $\theta$ is a vector, there may not be a simple or neat mathematical expression for {\it [R2]}. In some cases, a simultaneous CDs of a certain form for multiple parameters may not even exist (e.g., \citet{Xie2013a}). Fortunately, {\it [R2]} in Definition 1 only requires one set of confidence regions of all levels. %, to suffice in drawing valid inferential statements. 
In this manner, there are several options for defining multivariate CDs, e.g., \citet{Singh2007}; \citet{Xie2013a}; \citet[chap. 9 \& 15]{Schweder2016} and \citet{Liu2020}.

\begin{figure}
\centering
	\caption{{\it A graphical illustration of the inferential information contained in a CD as found in \citet{Xie2013a}. Examples pictured include point estimators (mode $\hat{\theta}$, median $M_n$, and mean $\bar{\theta}$), a $95\%$ CI, and a one-sided $p$-value.}}\label{fig:CD}
	\includegraphics[width=.8\linewidth]{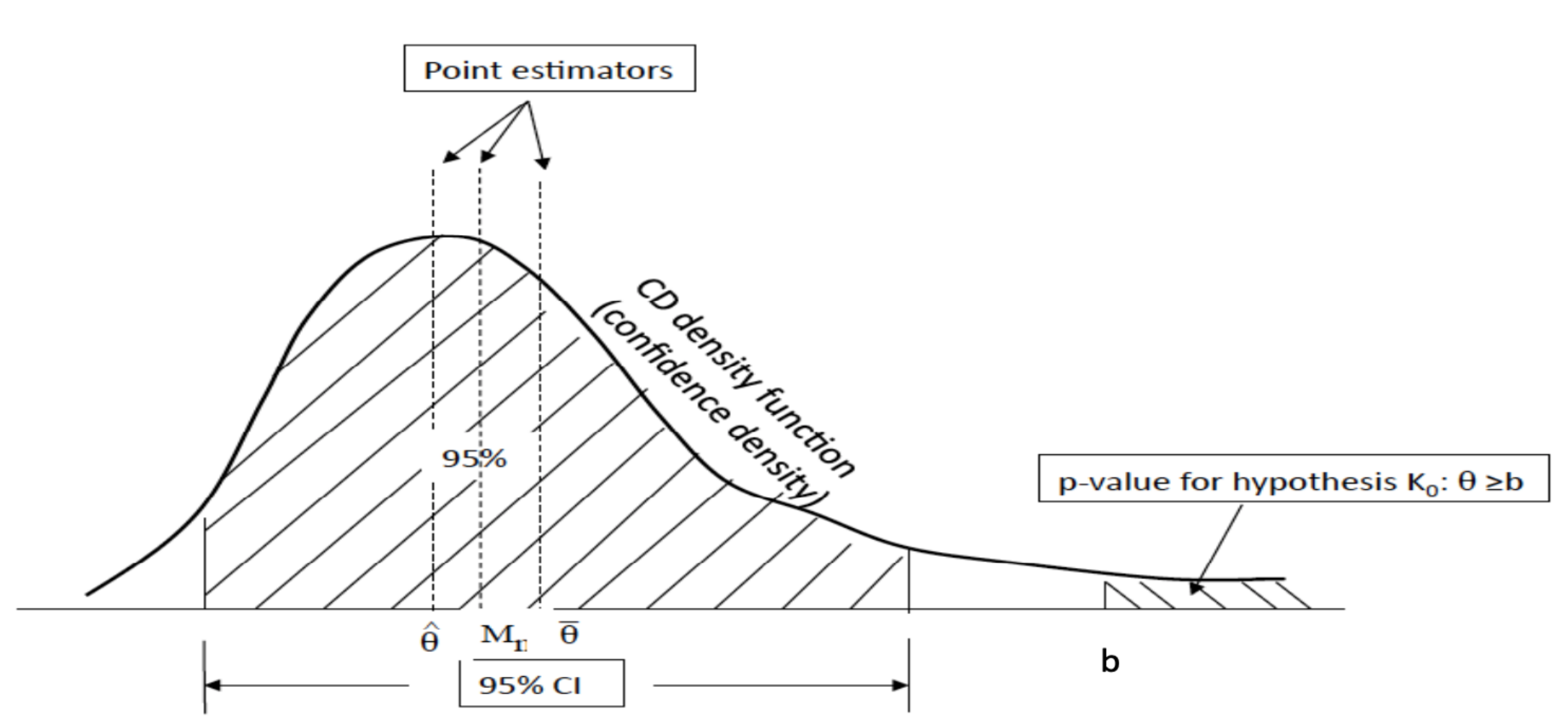}
\end{figure}

The idea of using a sample-dependent distribution function as an estimator of the target parameter is not new or unique to CD theory. A bootstrap distribution is sample-dependent and it is a distribution estimator for the unknown target parameter \cite[]{Efron1998}. In Bayesian inference, the posterior distribution is also a distribution estimator for the unknown target parameter and it too depends upon the observed sample. The benefit of using an entire function as an estimator, rather than an interval or a point, is that these distribution functions carry a greater wealth of information about the parameter of interest. A CD estimator, like a Bayesian posterior distribution, enables the statistician to draw inferential conclusion about an unknown parameter. For example, if %one is 
provided %with 
a CD for some parameter of interest, one can readily derive a point estimate, an interval estimate, and a $p$-value, among other features. This is illustrated in Figure \ref{fig:CD}. More details can be found in \citet{Singh2007, Xie2013a} and \citet[chap. 3]{Schweder2016}. 

The simple, illustrative example presented next will be revisited throughout the chapter. 

\begin{example}\label{ex:norm}%{\bf Inference for Gaussian Data with Confidence Distributions}
Suppose we observe an independent sample, $\by = (y_1, \dots, y_n)$, from a $N(\theta, 1)$ distribution. To estimate the unknown $\theta$ we may consider 1) a point estimate, e.g. $\bar{Y} = (1/n)\sum_{i=1}^{n}y_i$;  2) an interval estimate, e.g. $\left(\bar{Y}- 1.96/\sqrt{n}, \bar{Y}+ 1.96/\sqrt{n}\right)$ or 3) a distribution estimate, e.g. $H_n(\theta) = \Phi(\sqrt{n}(\theta - \bar{Y}))$, were $\Phi(\cdot$ is the cumulative distribution function of the standard normal distribution. This distribution estimate is %actually 
a CD for $\theta$. 

By itself, $H_n(\theta)$ provides information about $\theta$ including a point estimate, e.g. the center of the CD: $H_n^{-1}(1/2) = \bar{Y}$; an interval estimate, e.g., the interval formed by the upper and lower level-$(\alpha/2)$ CD quantiles: $\big(H_n^{-1}(\alpha / 2), H_n(1 - \alpha / 2)\big) = \big(\bar{Y} + \Phi^{-1}(\alpha/2)/\sqrt{n}, \bar{Y} + \Phi^{-1}(1-\alpha/2)/\sqrt{n}\big)$; and a $p$-value for the test $K_0: \theta \ge b$ versus $K_1: \theta < b$, e.g. the tail mass of the CD: $H_n([b, \infty)) = 1 - H_n(b) = 1-\Phi(\sqrt{n}(b - \bar{Y}))$.
\end{example}

Another important concept related to CD-based inference is that of a {\it CD-random variable} (henceforth CD-r.v.). For a given sample, a CD is a distribution function defined on the parameter space. We can construct a random variable (or vector), $\theta^*_{CD}$, on the parameter space such that, conditional upon the observed sample of data, $\theta^*_{CD}$ follows the CD. In the context of Example \ref{ex:norm}, we can simulate a CD-r.v. 
$\theta^*_{CD}$ by generating 
$\theta^*_{CD} \mid \bar{Y} = \bar y \sim N(\bar{y},1/n).$ This $\theta^*_{CD}$ can be understood as a ``CD-randomized'' estimator of $\theta$ (cf., \citealp{Xie2013a,Zhang2020}). 

\begin{definition}
Let ${\bf y} = (y_1, \dots, y_n)$ be an observed sample of data drawn from a distribution involving parameter $\theta$ and let $H_{n}(\theta)$ be a CD for $\theta$. Then, $\theta^{*}_{CD} | {\bf y} \sim H_{n}(\cdot),$ is referred to as a \emph{CD-random variable}. 
\end{definition}
%For an illustration see Figure 2 (a).

Defining the CD-r.v. by conditioning upon the data is similar to the conditioning required in bootstrap and  Bayesian inferences. Indeed, a CD-r.v. has many similarities to a bootstrap estimator or a random $\theta$ generated from a posterior distribution. 
We will explore these connections further in Sections~\ref{sec:scope} and \ref{sec:roots}. 

CD-based inference is incredibly flexible and can be presented in three forms: as a density function, a cumulative distribution function, or a {\it confidence curve}. If the target parameter $\theta$ is a scalar, a confidence curve is simply a mathematical representation of all two-sided confidence intervals for $\theta$ along every level of $\alpha$ on the vertical axis, i.e., $CV(\theta) = 2\min\{H_{n}(\theta), 1- H_{n}(\theta)\}$; e.g., \citet{Xie2013a};  \citet[chap. 1]{Schweder2016}. Also, see \citet{Liu2020} for discussions of confidence curves when the target parameter $\theta$ is a vector.
%{\color{red} Include more technical definition of CV here?}
In Example \ref{ex:norm}, a confidence density is $h_{n}(\theta) = (1/\sqrt{n})\phi(\sqrt{n}(\theta - \bar{Y})),$ where  $\phi(\cdot)$ is the density function of the standard normal distribution. Alternatively, a CD in the cumulative distribution function form is $H_{n}(\theta)=\Phi(\sqrt{n}(\theta-\bar{Y}))$ and a confidence curve is the function $CV(\theta) = 2 \min \{ \Phi(\sqrt{n}(\theta - \bar{Y}), 1-\Phi(\sqrt{n}(\theta - \bar{Y})\}.$ 
%, and $\Phi$ is the corresponding distribution function. 
Figure \ref{fig:pval} (a) shows 
% an image of 
a CD density, CD function, and confidence curve for the unknown location parameter using a random sample of size $n=12$ drawn from a $N(\theta_0 = 10,\sigma^2=1)$ distribution. 

In settings with nuisance parameters, a CD can provide an informative estimator for the parameter of interest (or the focus parameter). 
The Neyman-Scott example below is a slight modification of an example % found 
in \citet[\textsection 4.4]{Schweder2016}.

\begin{example}\label{ex:neyscott}
In the classical Neyman-Scott problem, we observe $2n$ independent pairs of data such that 
$$Y_{i,1},Y_{i,2} \sim N(\mu_i, \sigma^2), \text{ for }i=1,\dots,n;$$
i.e. each data pair is drawn from a distribution with the same variance $\sigma^2$ but with different mean $\mu_i$. % necessarily a common mean. 
The focus parameter is the common $\sigma^2$.  
Even with a large number of nuisance parameters, we still can construct a CD for $\sigma^2$ once we realize that 
$$
%\frac{\hat{\sigma}^2}{\sigma^2} = 
\frac{(1/n)\sum_{i=1}^{n}[(Y_{i,1}-\bar{Y}_i)^2 + (Y_{i,2}-\bar{Y}_i)^2]}{\sigma^2} \sim \frac{\chi^{2}_{(n)}}{n}$$
is free of the nuisance parameters $\mu_i$'s. 
Here, $\bar{Y}_i = (Y_{i,1} + Y_{i,2})/2.$
Specifically, let $C_{\nu}$ represent the cumulative distribution function for a $\chi^{2}$ random variable with $\nu$ degrees of freedom. 
A CD for $\sigma^2$ is then $H_{n}(\sigma^2) = 1 - C_{n}(2n\hat{\sigma}_{obs}^2/\sigma^2)$, where $\hat{\sigma}_{obs}^2 = (1/2n)\sum_{i=1}^{n}[(y_{i,1}-\bar{y}_i)^2 + (y_{i,2}-\bar{y}_i)^2]$.
The corresponding confidence density and confidence curve are
$h_n(\sigma^2) = [2n\hat{\sigma}_{obs}^2/ \sigma^4] c_n(2n\hat{\sigma}_{obs}^2/ \sigma^2)$ and $CV(\sigma^2)= 2 \min\{1 - C_{n}(2n\hat{\sigma}_{obs}^2/\sigma^2),  C_{n}(2n\hat{\sigma}_{obs}^2/\sigma^2) \}$, respectively, where $c_n(t) = C_n'(t)$ is the density function of the $\chi_n^2$ distribution. Figure \ref{fig:pval} (b) shows a CD density, CD function, and confidence curve for the unknown common variance from a sample of $n=100$ independent data pairs. In this particular example, the data was drawn from a $N(\mu_j, \sigma_0^2=1),$ for $j=1,\dots,n$ and the $\mu_j$'s are drawn from a U$(0,10)$ distribution.  
\end{example}

When the sample data are  drawn from a discrete distribution, most CD-related work in the % modern 
literature considers the case where the sample size is large and one may use an asymptotic CD, % is used; 
e.g., \citet{Singh2007, Xie2013a, Hannig2012}; \citet[chap 3.]{Schweder2016}. However, the concept of a CD is also useful in the case where the sample size is finite and the data are drawn from %data is limited in size and drawn from 
a discrete distribution. In these situations, 
%although exact CDs may not be attainable,
one may % instead
examine the difference between the discrete distribution estimator and the standard uniform distribution to get an idea of the under/over coverage of the confidence intervals derived from the CD. The definitions of lower and upper CDs are provided below for a scalar parameter $\theta \in \Theta$. Upper and lower CDs thus provide inferential statements in applications to discrete distributions with finite sample sizes. 

\begin{definition}%{[Upper and Lower CDs]} 
\label{def:u-l}
A function $H_{n}^{+}(\cdot) = H_{n}^{+}(\bY, \cdot)$ on ${\cal Y} \times \Theta \rightarrow [0,1]$ is said to be an \emph{upper CD} for~$\theta \in \Theta$~if \\
{[R1]} For each given $\by \in {\cal Y}$, $H_{n}^{+}(\by, \cdot)$ is a monotonic increasing function on $\Theta$ with values ranging within $(0,1)$; and \\ 
{[R2]} At the true parameter value $\theta=\theta_{0}$, $H_{n}^{+}(\theta_0) = H_{n}^{+}(\bY, \theta_0)$, as a function of the sample $\bY$, is stochastically less than or equal to a uniformly distributed random variable $U \sim U(0,1)$, i.e.,
\begin{equation}\label{eq:upperCD}
\mathbb{P}(H_{n}^{+}(\bY, \theta_0) \leq t) \geq t, \text{\,for all $t\in(0,1)$.}
\end{equation} 

A \emph{lower CD} $H_{n}^{-}(\cdot) = H_{n}(\bY, \cdot)$ for parameter $\theta$ can be defined similarly, but with (\ref{eq:upperCD}) replaced by $\mathbb{P}(H_{n}^{-}(Y, \theta_0)\leq t) \leq t$ for all $t \in (0,1)$.
\end{definition}

Sometimes, even if condition {\it [R1]} in Definition~\ref{def:u-l}  does not hold, $H_{n}^{+}(\cdot)$ and $H_{n}^{-}(\cdot)$ are still referred to as upper and lower CDs, respectively. This is because regardless of whether or not the monotonicity condition holds, the stochastic dominance inequalities in the definition of upper and lower CDs ensure that for any $\alpha \in (0,1)$ 
\[\mathbb{P}\left(\theta_0 \in \{\theta : H_{n}^{+}(\bY, \theta) \leq \alpha \}\right) \geq \alpha \quad \text{and}\quad 
\mathbb{P}\left(\theta_0 \in \{H_{n}^{-}(\bY,\theta)\leq \alpha \} \right) \leq \alpha.\]
Thus, a level $(1-\alpha)$ confidence interval (or region) $\{\theta : H_{n}^{+}(\by,\theta) \leq 1-\alpha\}$ or $\{\theta : H_{n}^{-}(\by, \theta) \geq \alpha\}$ has guaranteed coverage rate of at least $(1-\alpha)$. Without condition {\it [R1]} however, $H_{n}^{+}(\cdot)$ and $H_{n}^{-}(\cdot)$ may not be distribution functions and confidence intervals (or sets) may no longer follow the nesting condition, e.g., \citet{Blaker2000, Xie2013a}. This means that a level $(1-\alpha)$ confidence set, $C_{1-\alpha}$, may not necessarily be contained in a corresponding level $(1-\alpha')$ confidence set, $C_{1-\alpha'}$, for some $ \alpha' < \alpha$.
This is a departure from the classical definition of a CD found in \citet{Cox1958}, where a CD is obtained by inverting the bounds of nested confidence intervals at every $\alpha$ level. 

The following is an example of an upper and lower CD for the parameter in a discrete distribution. For more details on this example, see \citet{Hannig2012}. A more complex example can be found in \citet{Luo2020}, which links Fisher's sharp null test in causal inference with the concepts of lower and upper CDs and their corresponding (extended) confidence curves.

\begin{example}%{\bf Upper and Lower CDs from a Binomial Sample}
Suppose we observe $y$ from a Binomial$(n, p_0)$ distribution. Let $H_n(p,y) = Pr(Y > y) = \sum_{y <         k \leq n_i} \binom{n}{k}p^k (1-p)^{n-k}$. It can be shown that $Pr(H_n(p_0, Y)\leq t)\geq t$ and $P(H_n(p_0, Y-1)\leq t)\leq t$. Thus, $H_{n}^{+}(p,y)=H_n(p,y)$, and $H^{-}_{n}(p,y)=H_n(p,y-1)$ are lower and upper CDs for the true success rate $p_0$. The half-corrected CD used in the literature (e.g., \citealt[\textsection 3.8]{Schweder2002}),
$$\frac{H^{-}_{n}(p,y) + H^{+}_{n}(p,y)}{2} = \sum_{y < k \leq n} \binom{n}{k}p^{k}(1-p)^{n-k} + \frac{1}{2} \binom{n}{k} p^{y} (1-p)^{n-y},$$
is an aCD. In the finite sample setting, we do not use the half-corrected CD, and
a level $(1-\alpha)$ confidence interval that guarantees the coverage rate is
$\{\theta : H_{n}^{+}(\by,\theta) \leq 1-
\frac\alpha2, H_{n}^{-}(\by, \theta) \geq \frac\alpha2\}$. 
\end{example}

\subsection{Construction with different inferential methods}\label{sec:construct}
The concept of a CD includes a broad class of common inferential methods, e.g., \citet{Xie2013a}. Among this class of methods are bootstrap distributions, (normalized) likelihood functions, (normalized) empirical likelihood functions, $p$-value functions, fiducial distributions, Bayesian posterior distributions, and more. As summarized in \citet{Cox2013}, the usefulness of inference from a CD perspective lies in its ability to ``provide simple and interpretable summaries of what can reasonably be learned from data (and an assumed model)''. Regardless of whether the setting is parametric or nonparametric, normal or non-normal, exact or asymptotic, CD inference is possible as long as one can create confidence intervals (or regions) of all levels for the parameter of interest. This fact illustrates the appeal of CD-based inference as uniting different inferential paradigms. Before moving on to the next section, we present a CD framework based on likelihood inference, and on hypothesis testing, with a few relevant examples.

Likelihood-based approaches and their extensions are arguably the most commonly used methods to answer questions of statistical inference. Asymptotic, large sample arguments are frequently used to quantify the uncertainty in our point estimators from likelihood-based approaches. The very same consequences of the central limit theorem also provide a connection between the likelihood function and a CD, provided ${\int L(\bm \theta \mid data)d\bm\theta} < \infty$, where $L(\bm \theta \mid data)$ is either the likelihood function or a generalized likelihood function. 
For instance, under some regularity conditions, \citet{Fraser1984} proved that a normalized likelihood function is an aCD and \citet{Singh2007} showed that a normalized profile likelihood function is also an aCD. The inferential conclusions based on these CDs asymptotically match the first-order conclusions based on the likelihood or profile likelihood functions. Higher-order asymptotic developments relating likelihood inference to CDs include, for instance,  \citet{Reid2010,  Reid2015, Pierce:Bellio2017}, among others. These higher-order asymptotic developments can also be used to derive more accurate CDs, e.g., \citet{Xie2013a}.

\addtocounter{example}{-3}
\begin{example}[continued]
The likelihood function is $L(\theta \mid \by) \propto  \prod_{i} f(y_i \mid \theta) \propto e^{-\frac{1}{2}\sum(y_i-\theta)^2 } \propto e^{-\frac{n}{2}(\bar{y}-\theta)^2 }.$ Normalizing $L(\theta \mid \by)$ with respect to $\theta$, we get 
$L(\theta \mid {\bf y})/\int L(\theta \mid {\bf y})d\theta =  \sqrt{n} \phi\left(\sqrt{n}(\theta-\bar{y}) \right)$
which is exactly a CD density function.
Furthermore, since we are only dividing by a constant when normalizing (standardizing) the likelihood function, the mode of the CD is the same as that of the likelihood function. Thus, inferential conclusions drawn from the likelihood function $L(\theta \mid \by)$ are exactly the same as the inferential conclusions drawn from the CD $H_{n}(\theta) =
\Phi(\sqrt{n}(\theta-\bar{Y}))$.
\end{example}\label{ex:like}  

There may be instances in which we are interested in a scalar parameter in a model with a vector of parameters. In this situation, it is common to use a profile likelihood for inference about the focus parameter, and the profile likelihood can also be used to derive a CD for the focus parameter; e.g.,  \citet{Singh2007}. Example \ref{ex:profile_like} below outlines one such case where a CD for the shape parameter of a Gamma distribution is derived from the profile likelihood function.

\addtocounter{example}{+2}
\begin{example}\label{ex:profile_like}
Suppose we observe a random sample of size $n$, say $\by = y_1, \ldots, y_n$, from a $Gamma(\theta,\beta)$ distribution, where the shape parameter, $\theta > 0$, is the focus parameter and the rate parameter, $\beta >0$, is a nuisance parameter. The likelihood function is $L(\theta, \beta \mid \by) = \prod_{i}[\beta^{\theta}/\Gamma(\theta)] y_i^{(\theta-1)}e^{-\beta y_i},$ where $\Gamma(x) = \int_{0}^{\infty}z^{(x-1)}e^{-z}dz$. The profile likelihood is found by substituting $\hat{\beta}(\theta)= \theta/ \bar{y}$ for $\beta$ in $L(\theta, {\beta} \mid \by)$. This can be normalized with respect to the parameter of interest thereby deriving a CD density for $\theta$  
$$h_n(\theta) \propto \frac{1}{\Gamma(\theta)^n} \left( \frac{\theta}{\bar{y}} \right)^{n\theta} \left(\prod_{i}y_i\right)^{(\theta-1)}e^{- n \theta}.$$
The corresponding CD, $H_n(\theta)= \frac{\int_0^\theta h(s) ds}{\int_0^\infty h(s) ds}$, can be obtained numerically.
\end{example}% 

Likelihood-based inference represents only one particular generative method for obtaining a CD since CDs can be obtained in many other cases without likelihood functions. In the example below, we consider an hypothesis testing approach, another common inferential method intimately linked to CDs. As with likelihood-based inference, hypothesis tests represent only one of many other generative methods by which a CD may be derived.  For simplicity, we consider only the case that the focus parameter $\theta$ is a scalar; In the case when $\theta$ is a vector, an example connecting CD with a testing problem can be found in  \citet{Liu2020}.  

Consider a one-sided hypothesis testing problem $K_0: \theta \leq  \theta_0$ versus $K_1: \theta >  \theta_0$ or a two-sided test $K_0: \theta =  \theta_0$ versus $K_1: \theta \not =  \theta_0$.   Denote by $p_n= p_n(\theta_0)$ the $p$-value from a testing method; this $p$-value depends on both the observed sample and the tested value of the parameter, $\theta_0$. As $\theta_0$ varies in the parameter space $\Theta$, $p_n = p_n(\theta_0)$ is a function on the parameter space called a {\it $p$-value function} or a {\it significance function}; see, e.g., \citet{Fraser:1991}.  Often, $p_n(\theta_0)$ (as a function of the random sample) follows a standard uniform distribution under the the null hypothesis. As function of $\theta_0$, $p_n(\theta_0)$ is typically monotonically increasing in the one-sided hypothesis test setting, and first increases and then decreases in the case of a two-sided hypothesis test.  \citet{Singh2007} and \citet{Xie2013a} show that the $p$-value function from the one-sided test corresponds to a regular CD function and the $p$-value function from a two-sided test corresponds to a confidence curve.

\begin{figure}
\centering
	\caption{{\it Each row displays a CD density, a CD curve, and a confidence curve (CV) for various examples in this chapter. Plot (a) is derived from Example \ref{ex:norm} with a random sample of size $n=12$. Plot (b) is derived from Example \ref{ex:neyscott} for $n=100$ observations of independent, paired data points. Plot (c) is derived from Example \ref{ex:mannwhit} with two independent samples of size $n_1=10$ and $n_2=9$. In each plot, the values of the horizontal axis at intersections between the gray dashed line(s) and the solid (CD) curves are the lower and upper bounds of the corresponding $95\%$ confidence interval.  }}\label{fig:pval}
	\begin{subfigure}{.9\linewidth}
	    %\hspace{2mm}
	    \centering
	    \includegraphics[width = 0.82\linewidth]{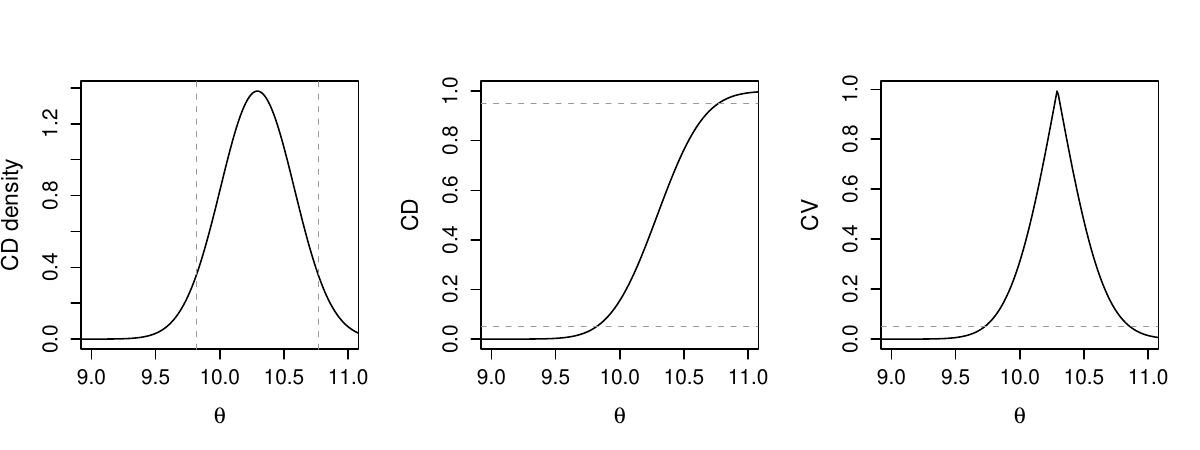}
   	    \caption*{(a) CD-inference for the location parameter of a Normal random sample, Example \ref{ex:norm}.}
	    \end{subfigure} 
	 \begin{subfigure}{.9\linewidth}
	    \hspace{2mm}
  	    \centering
    	\includegraphics[width = 0.82\linewidth]{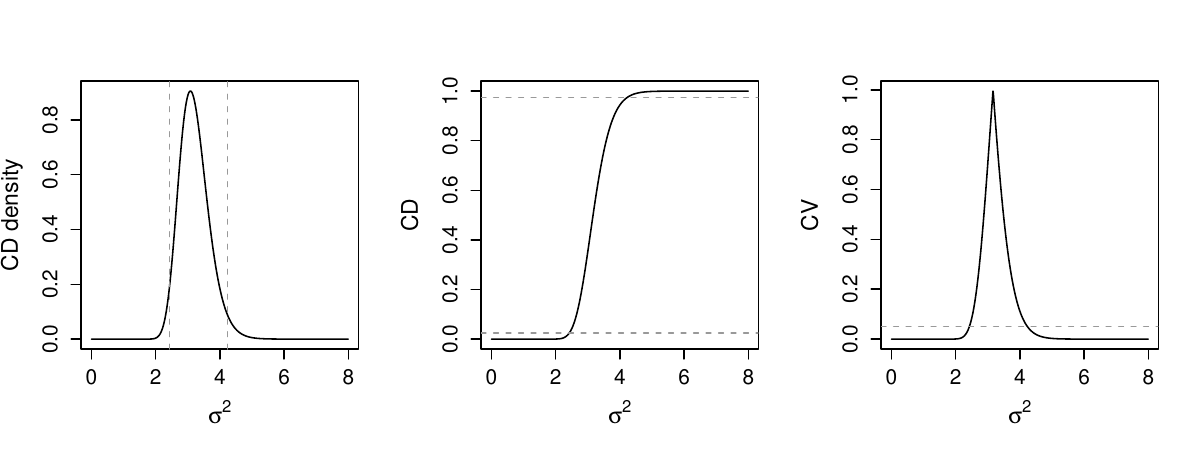}
    	\caption*{(b) CD-inference for the common variance in the Neyman-Scott setting, Example \ref{ex:neyscott}.} 
    	\end{subfigure}
	\begin{subfigure}{.9\linewidth}
	    \hspace{2mm}
   	    \centering
    	\includegraphics[width = 0.82\linewidth]{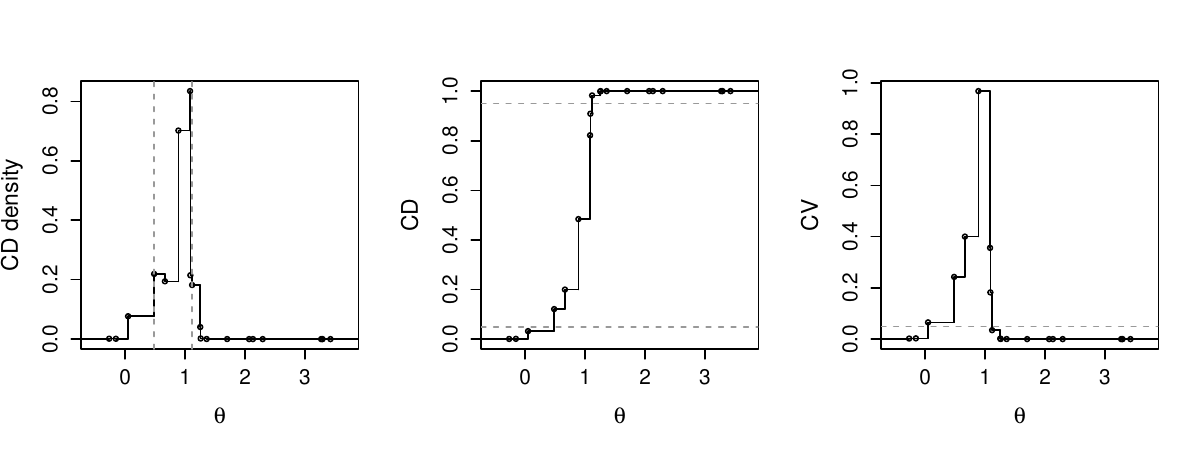}
    	\caption*{(c) CD-inference for the location shift from the Mann-Whitney setting, Example \ref{ex:mannwhit}.} 
    	\end{subfigure}
\end{figure}

\addtocounter{example}{-4}
\begin{example}[continued]\label{ex:pval1} 
For a one-sided hypothesis test of $\theta$, $K_{0}: \theta \leq  \theta_0$ versus $K_{1}:\theta > \theta_0$, the $p$-value is calculated by 
$q_n(\theta_0) = %\mathbb{P}(\bar{Y} > \bar{Y}) = 
\Phi(\sqrt{n}(\theta_{0}-\bar{y})).$ 
As $\theta_0$ varies in $\Theta$, $q_n(\theta_0)$ is the cumulative distribution function of a  $N(\bar{y},1/n)$ distribution and is therefore a CD for $\theta$. For the two-sided hypothesis test $K_{0}: \theta = \theta_0$ versus $K_{1}:\theta \neq \theta_0$, the $p$-value is calculated by 
$2 \min\{q_n(\theta_0), 1-q_n(\theta_0) \} = 2\min\{\Phi(\sqrt{n}(\theta_0-\bar{y})), \Phi(\sqrt{n}(\bar{y}-\theta_0)) \},$
which is a confidence curve for $\theta$. See Figure \ref{fig:pval} (a).  
\end{example}

The next example considers the Mann-Whitney test for determining whether two independent samples are drawn from the same population. For a careful discussion on setting up the hypotheses for a Mann-Whitney test, we refer the reader to \citet{Divine:etal2018}. 

\addtocounter{example}{+3}
\begin{example}\label{ex:mannwhit} 
Consider two independent samples $\bx = (x_1, \dots, x_{n_1})$ and $\by=(y_1, \dots, y_{n_2})$ each drawn from an unknown continuous distribution (say, $F(\cdot)$ and $G(\cdot)$, respectively) that differ in location only. Without loss of generality, suppose $n_1 > n_2$ and let $n_1 + n_2 =n$. In this case, the Mann-Whitney test for a difference in the population distributions is equivalent to testing the null that $K_0: F(t) = G(t), \text{ for all $t$}$ versus the alternative $K_1: F(t) = G(t-\theta), \text{ for every $t$ and for some $\theta \neq 0$}.$ The Mann-Whitney test statistic is $U_{\theta} = n_1n_2 +\frac{n_2(n_2+1)}{2} - R_\theta,$ where $R_\theta$ is the sum of the ranks of $\bX$ among the set $\{X_1, \dots, X_{n_1}, Y_1 + \theta, \dots, Y_{n_2}+\theta \}$. Let $A_{\theta_0}(t) = P(U_{\theta} \leq t \mid \theta=\theta_0)$ be the distribution of $U_{\theta}$ under the null hypothesis $K_0$. Then, the $p$-value for this test is calculated by 
$2\min\{A_{\theta}(u_{\theta}), 1-A_{\theta}(u_{\theta})\},$
where $u_{\theta}$ is the value of $U_{\theta}$ evaluated at $\bx$ and $\by$. This generates a confidence curve for $\theta$. Also, we can show that $H_n(\theta) = A_{\theta}(u_{\theta})$ is a CD function, from which a CD density can be readily derived. Figure \ref{fig:pval} (c) shows an image of a CD density, CD function, and confidence curve for the unknown location shift where one sample $(n_1=10)$ is drawn from a t-distribution with $\nu=5$ degrees of freedom, and the other $(n_2=9)$ is independently drawn from a t-distribution also with $\nu = 5$ degrees of freedom but also with a non-centrality parameter $\theta_0=1$.
\end{example}

\section{Using the CD framework to unify and combine inference across the BFF paradigms}\label{sec:scope}

A distinguishing feature of CDs is that this type of estimation provides a framework under which results across different inferential paradigms can be directly compared or even combined, as will be demonstrated in Example \ref{ex:bivar}. One may better understand how this is possible by recognizing that CD theory extends several practical considerations behind point estimation to the class of distribution estimators. To make this clear, let us first consider the context of statistical point estimation. 

Point estimation as a concept itself is inherently arbitrary. An estimator must be a sample dependent function that produces a point in the parameter space. The rich mathematical literature on particular methods of point estimation and their comparison somewhat paradoxically thrives through setting restrictions. For instance, one may be interested in estimators that behave according to certain impartiality conditions such as unbiasedness or equivariance. Still more desirable properties may include consistency or efficiency. These traits are ubiquitously recognized as favorable among statisticians working within any of the Bayesian, fiducial, or frequentist paradigms, but are nothing more than performance-based descriptors of a potential estimator. That is, the pursuit of statistical estimators from this perspective reflects a behavioral reasoning process.
A point estimator defined by behavioral reasoning, however, does not necessarily include instructions for obtaining the estimate. For example, the definition of the consistent estimator does not provide any direction on how the consistent estimate can be obtained. Instead, a consistent estimate is often obtained in collaboration with a procedural-based approach.

Examples of procedurally defined estimators include the maximum likelihood estimator, moment estimator, Bayes estimator, and others. Most of these procedurally defined estimators are rooted in a certain type of inductive reasoning often supported by decision theory, whereby a risk function is minimized. These procedural methods are frequently applied in the search for desirable behaviorally-defined estimators, but they generally do not provide universal guarantees about the performance of the resulting estimator. Instead, the starting point afforded by these various procedures provides some orientation in a search that may otherwise be quite arbitrary. The reasons for choosing a particular procedure-based approach are varied and highly problem-dependent but what is universally appealing is that the resulting estimators behave according to some performance-based quality.  

Now, we may broaden our perspective to also consider distribution estimators. A CD is defined according to two performance-based descriptors; recall {\it [R1]} and {\it [R2]} from Definition \ref{def:CD}. The requirement {\it [R2]} describes the performance of a distribution estimator that covers the parameter of interest with at least a specified frequency in repeated, independent experiments. This so-called relative frequency principle has become a (nearly) universally appealing standard for useful statistical inference regardless of the inferential framework. Thus one may consider different inferential methods as a starting point in the search for a well-behaved distribution estimator. For example, the Bayesian method derives a posterior distribution which can be thought of as a distribution estimator for the unknown parameter. As mentioned earlier, within a frequentist framework one may derive a $p$-value function or a normalized likelihood function to obtain a distribution estimator. The fiducial distribution is another distribution estimator that can be derived in many settings through the method of generalized fiducial inference \citep{Hannig2016}. There are innumerable ways in which one may derive a distribution estimator, but a CD inferential framework suggests restricting this endeavour according to a tried-and-true (frequency) property that can be easily explained and communicated to the broader scientific community.  
% \ST{property that has advanced modern applications of statistics in medicine and technology and more.} 

Table \ref{tab:behaviorist} compares these different motivating (or restrictive) qualities and methods across the broad landscape of statistical estimation. For instance, a consistent estimator is defined by requiring a certain behavior, and it is often obtained via a procedural-based estimator, such as an MLE approach or some other method. On its own, the procedural-based MLE, is obtained through steps devised in likelihood inference. This estimator is often consistent; however, this is not guaranteed and the statistician must verify its performance in different contexts. Similarly, a CD is a behaviorally defined estimator and this definition does not provide any direction on how it may be obtained. Practically, a CD can often be obtained via a procedural-based approach, for example, from a p-value function via hypothesis testing, or a bootstrap distribution via resampling, or a fiducial distribution via inversion, or a posterior distribution via Bayes formula, and so forth. A fiducial distribution and Bayesian posterior, on the other hand, are procedurally defined estimators supported by inductive reasoning. A fiducial distribution and Bayesian posterior are often, but not always, a CD or aCD, and one must verify the requirements hold in different contexts.

\begin{table}[hb]
\centering
\caption{A table depicting a range of ways in which a statistical estimator may be defined.}
\begin{tabular}{rllll}
\hline \hline
 \textbf{Estimator} && \multicolumn{2}{l}{\textbf{Definition}} \\ \hline 
\textbf{Type} & & \textbf{Behavior}    && \textbf{Procedure}\\
 & & {(Desired performance)}    && {(Inductive reasoning and/or operational decision)}\\\hline
Point   & & Consistent estimator    && Maximum likelihood estimator, M-estimation, \\
& &  &&
 Moment estimator, Bayes estimator  \\ \hline
Distribution & & Confidence distribution &&  Bootstrap distribution, Fiducial distribution, \\  
& & &&(normalized) likelihood, $p$-value function, \\ & & && Bayesian posterior 
 \\ \hline
\end{tabular}
\label{tab:behaviorist}
\end{table}

By approaching the problem of statistical inference from a more general perspective targeting well-behaved distribution estimators, CD theory provides a common platform to compare and even combine inferential outcomes across different paradigms in ways that would otherwise not be possible. The example below, inspired by \citet{CuiandXie},  presents a CD-based meta-analysis approach that combines the results from four independent studies across three different inferential paradigms. The unique feature of this particular example is that, in each of the four independent studies, one particular inferential method, either frequentist, Bayesian, or fiducial, was adopted to initially analyze the data and the results of these initial analyses provide only summary statistics for use in a CD-based meta-analysis that combines all results. We refer the interested reader to \citet{Xie2011} and \citet{Singh2005} for more detailed developments on CD-based meta-analysis and CD combining methodologies.
 
\begin{figure}
\centering
	\caption{{\it A visual comparison of one instance of the meta-analysis problem from Example \ref{ex:bivar}. The four individual inferential functions for $\rho$ are derived from four independent studies across three inferential paradigms and the combined CD approach (bottom row) incorporates information from all four studies. The target value of $\rho_0 = 0.723$ is indicated by the dashed vertical
	line.}}\label{fig:bivar}
	{\includegraphics[width=0.93\linewidth, height = 8cm]{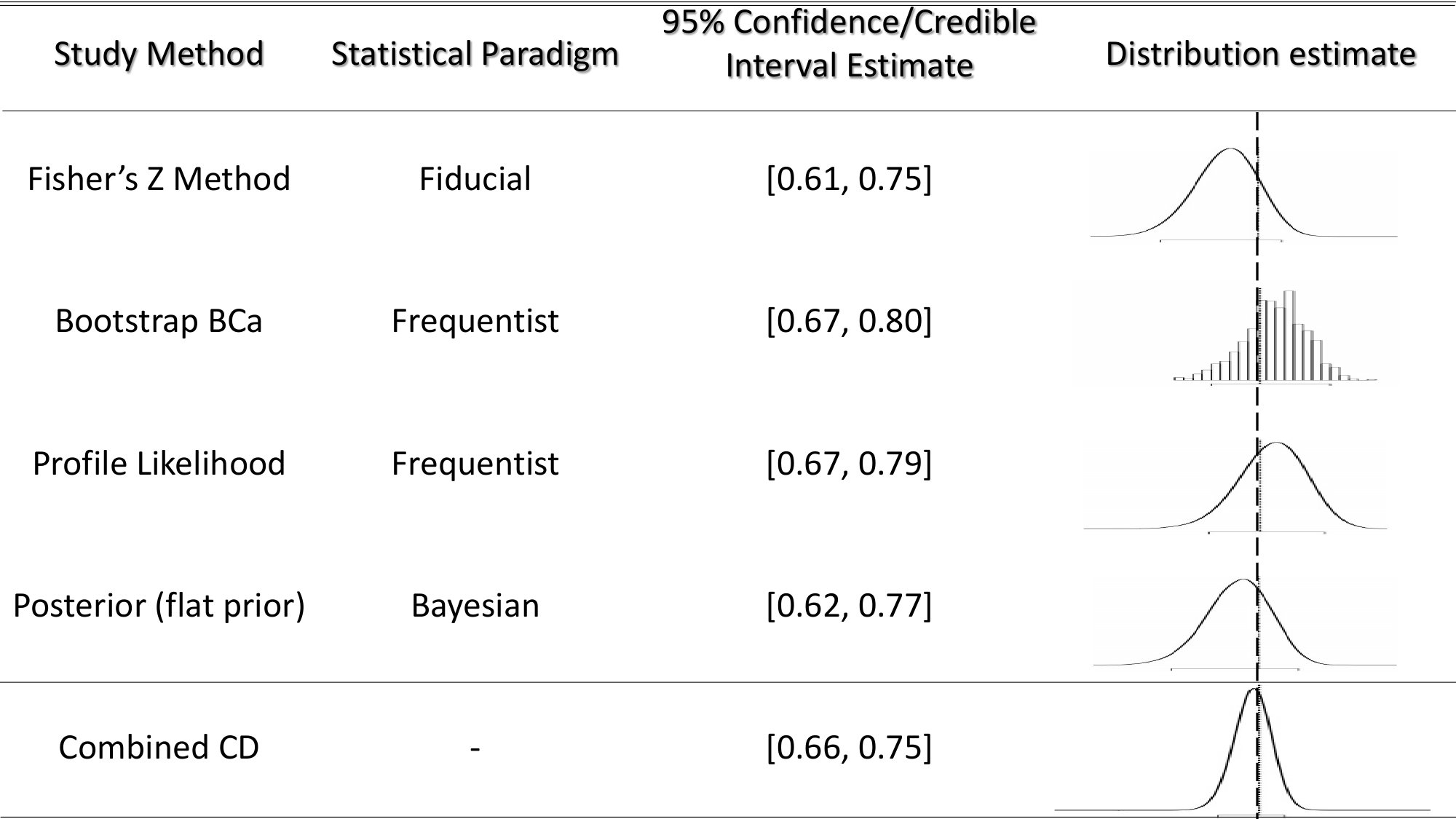}}  
\end{figure}

\begin{example}\label{ex:bivar}
Suppose we have $k=4$ independent studies in a typical meta-analysis setup where each study collects $n$ independent observations of some count data before and after a new treatment, $\by^{(k)} = (\by_{1}^{(k)}, \by_{2}^{(k)})^{T}$. For each individual $i$ in the $k$th study, the paired observations are assumed to follow a bivariate normal distribution
\begin{equation}\label{eq:cd4}
{\bf y}_i^{(k)} = \begin{pmatrix}
y_{i1}^{(k)} \\ y_{i2}^{(k)}
\end{pmatrix} \sim N\left( 
\begin{pmatrix}
\mu_1 \\ \mu_2
\end{pmatrix}, 
\begin{pmatrix}
\sigma_1^2 & \rho \sigma_1 \sigma_2 \\ 
\rho \sigma_1 \sigma_2 & \sigma_2^2 
\end{pmatrix}
\right), \,\, \hbox{
for $i = 1, \ldots, n$ and $k = 1, 2,3,4$}.
\end{equation}
Consider the correlation cofficient, $\rho$, as the parameter of interest and $(\mu_1, \mu_2, \sigma_1^2, \sigma_2^2)$ as nuisance parameters. 

Say study $k=1$ uses Fisher’s $Z$ method \citep{Fisher1915}
and, by inverting the distribution of an asymptotic pivot statistic to obtain a distribution estimator \citep{Singh2007}, we treat it as a fiducial approach. Study $k=2$ uses a bias-corrected and accelerated (BCa) bootstrap method \citep{DiCiccio1996, Davison:Hinkley1997, Canty:Ripley2019}
and study $k=3$ uses a profile likelihood approach.  %\citep{Li:etal2018}.
Hence, each of these two studies can produce a CD and draw inferential conclusions within a frequentist framework.
Study $k=4$ produces a Bayesian posterior distribution based on a uniform prior. The combined CD-approach draws from each of these $k=4$ studies by treating all four distribution estimators as CDs, and produces a fifth distribution estimator. We study the performance of the fifth distribution estimator and compare it with the other four distribution estimators.

In particular,  
$H_1(\rho) 
= H_1(\rho, \by^{(1)}) = 1 - \Phi\left[\frac{\sqrt{n-3}}2( \log\frac{1+r}{1-r} - \log\frac{1+\rho}{1-\rho})\right]$ 
is the distribution estimator derived from the Fisher's Z test statistic, where $r$ is the sample correlation coefficient; \cite[cf, e.g.,][]{Singh2007}. The corresponding CD density is $h_1(\rho) =\frac{\sqrt{n-3}}{2(1 - \rho^2)}\phi\left[\frac{\sqrt{n-3}}2( \log\frac{1+r}{1-r} - \log\frac{1+\rho}{1-\rho})\right] $.
Let $h_{2}(\rho) = h_2(\rho,  \by^{(2)})$ be the bootstrap (density) distribution using the bootstrap BC method. Similarly, let $h_{3}(\rho) = h_{3}(\rho, \by^{(3)})$ be the normalized profile likelihood function and $p_4(\rho) = p_4(\rho |\by^{(4)})$ be the posterior distribution. The key to the combined CD approach is realizing that each of $h_i(\rho)$, $i = 1, 
\ldots, 4$ are individual CD densities that summarize the inferential information available for $\rho$ from each individual study. Thus, a combined CD that incorporates the information from all four studies can be expressed explicitly as
\begin{equation*}
    H^{(c)}(\rho) = G_c\left( g_c\bigg(\int_{-1}^\rho h_1(s) ds,  \int_{-1}^\rho h_2(s) d s,  \int_{-1}^\rho h_3(s) d s, \int_{-1}^\rho p_4(s) d s \bigg)\right), 
\end{equation*}
where $g_c(u_1, u_2, u_3, u_4)$ is a given monotonic mapping function from the quartic cube to the real line
%, $(0,1)^4 \to \reals$, 
and $G_c(t) = P\left(g_c(U_1, U_2, U_3, U_4) \leq t\right)$ with $(U_1, \ldots, U_4)$ being IID $U(0,1)$ random variables.  
In the numerical study below, we have used $g_c(u_1, \ldots, u_4) = DE^{-1}(u_1) + \ldots + DE^{-1}(u_4)$,
where $DE^{-1}(\cdot)$ is the inverse cumulative distribution function of the standard Laplace distribution. With this choice of $g_c$, the combining approach is Bahadur efficient \citep{Singh2005, Singh2007}.

Figure \ref{fig:bivar} illustrates the similarities of the individual inferential functions $h_1(\rho)$, $h_{2}(\rho)$, $h_{3}(\rho)$, and $p_{4}(\rho)$ and the combined CD density, $h^{(c)}(\rho) = \frac{\partial}{\partial\rho} H^{(c)}(\rho)$ as computed by the {\it gmeta} function in the R package of the same name \citep{Yang2017}. This figure is based on four independent samples drawn from a bivariate normal distribution with $n = 200$ and true parameter values $\mu_1 = 3.288,$ $\mu_2 = 4.093$, $\sigma_1^2 = 0.657$, $\sigma_2^2 = 1.346$, and $\rho_0 = 0.723$. Table \ref{tab:my-table} summarizes the relative performance of each of these five different approaches based on a simulation study with $1000$ repeated samples from the same bivariate normal distribution. We see that the combined approach still maintains the nominal coverage level though the interval width of this approach is roughly half that of the other four individual approaches. This result is unsurprising since theoretically each study provides a $\frac1{\sqrt{n}}$-confidence interval and the combined method provides a $\frac1{2\sqrt{n}}$-confidence interval, noting that the total sample size of the combined data is $4n$. This example demonstrates how inference across different paradigms can be combined through a CD approach. 
%we can combine inference across different paradigms and provides a suggested implementation.
\end{example}
 
\begin{table}
\begin{center}
\caption{{\it Coverage results from $1000$ simulated data sets for each of the five different inferential methods mentioned in Example \ref{ex:bivar}. The last row represents a CD approach that combines individual-study inference from the four independent studies.} }\label{tab:bivar}
\label{tab:my-table}
\begin{tabular}{lll}
\hline\hline
\multicolumn{1}{c}{\textbf{Method}} & \multicolumn{2}{c}{\textbf{$95\%$ CI}}                                                \\ 
\textbf{}                           & \multicolumn{1}{c}{\textbf{Coverage}} & \multicolumn{1}{c}{\textbf{Mean length (sd)}} \\ \hline 
Fisher's $Z$ method                 & $0.939$                               & $0.132$ $(0.014)$                             \\
Bootstrap BC$_a$                    & $0.931$                               & $0.132$ $(0.018)$                             \\
Profile likelihood                  & $0.952$                               & $0.130$ $(0.013)$                           \\
Bayes (uniform prior)               & $0.951$                               & $0.139$ $(0.015)$                            \\ \hline 
Combined CD                         & $0.944$      & $0.064$ $(0.006)$                            \\\hline
\end{tabular}
\end{center} 
\end{table}
 
%---------------------------------
\section{BFF Union: Randomness matching and parameter duality}\label{sec:roots}  

Statistical inference is primarily concerned with the quantification of uncertainty inherent in sampling data from a %larger
population. Since we observe only a single copy of the data, assessing this uncertainty %to assess the uncertainty we 
typically requires an assumed statistical model
%accurately describes the behavior of the population at large. 
% Under the statistical model assumption, 
under which the observed data are considered to be a particular realization of a random sample set. 
% In some literature, the uncertainty associated with random sampling is referred to as ``empirical varibility''; cf., \citet{Reid2015}.
%The uncertainty associated with the random sample, referred to as ``empirical varibility'' in \citet{Reid2015}, is the target of most, if not all, statistical inference approaches.
Sometimes an inference problem is tractable mathematically 
% and in this case
%so that we can assess this uncertainty 
and assessments of empirical variability are possible through an explicit mathematical formulation.
% that may or may not relay on asymptotic, large sample size, arguments. %either exactly or asymptotically. 
In more complicated situations, incorporating additional simulation-driven methods is necessary to assess the empirical variability. This section focuses on the role simulation-driven methods play in each of the three BFF paradigms. 
 
We specifically highlight a common trait that matches two types of randomness: the artificial randomness induced by a simulation mechanism and the empirical randomness inherited from the sampling procedure under the assumed model. 
Following the exploration of this matching condition, we revisit the discussion as to whether or not a parameter should be viewed as random or fixed or if perhaps both perspectives are applicable. We try to avoid a philosophical discussion of {\it aleatory} versus {\it epistemic} probability %. Rather, we will directly 
and instead focus on matching the probability measures induced by artificial simulation (i.e., the one that governs the artificial randomness) and inherited from the assumed model (i.e., the one that describes the empirical randomness). A brief comment on how these  probability measures may %or may not 
relate to aleatory and epistemic definitions of probability is provided in Section \ref{sec:discussion}.

For ease of presentation, this section uses the notation %we define a matching notation 
$\widesim{m}$  to relate two real-valued random quantities of the same dimension, say, $W_n$ and $V_n$ of dimension $p$, that are defined on two different probability spaces, one associated with the artificial Monte-Carlo sample space and the other from the standard sample space. Specifically, the expression $W_n \widesim{m} V_n$ refers to either an {\it exact} matching such that 
\begin{equation}
\label{eq:Prob-exact}
\mathbb{P}_A(W_n \in C | data) = \mathbb{P}(V_n \in C),
\end{equation}
in probability, or an {\it approximate} matching (often reliant upon $n \to \infty$) such that 
\begin{equation}
\label{eq:Prob-asy}
\big| \mathbb{P}_A(W_n \in C | data) - \mathbb{P}(V_n \in C)\big| \to 0, 
\end{equation}
in probability, for any Borel set, $C$, in the Borel $\sigma$-field on ${\bf R}^p$.
% , within the respective probability space. 
Here, the probability statement $\mathbb{P}_A(\cdot|data)$ is defined according to the artificial randomness governed by the Monte-Carlo simulation procedure, conditioned upon an observed data set, and $\mathbb{P}(\cdot)$ is defined according to the usual empirical randomness as determined by the sampling procedure under the assumed model. 
The uncertainty described by $\mathbb{P}(\cdot)$ is typically the target of interest for frequentist inference. As we will see next, most literature exploring such matching conditions rely on asymptotic arguments; however, our consideration includes the exact matching of (\ref{eq:Prob-exact}) as well.
 
\subsection{Monte-Carlo simulation and a matching condition for statistical inference}\label{sec:inversion}

Monte-Carlo simulation methods have proven to be incredibly useful and effective tools for statistical inference for which there exists a rich literature,  e.g., \citet{EfroTibs93,fishman1996,glasserman2003}, 
and references therein.
One popular approach is to create many copies of artificial (or ``fake'') data using a Monte-Carlo procedure, and then to use these fake data to assess statistical uncertainty. 
This type of schema appears throughout the BFF paradigms including, as will be explored in this section, the bootstrap sampling method, generalized fiducial inference (GFI), and approximate Bayesian computing (ABC) 
\citep[cf., e.g.,][]{EfroTibs93,ernst2004, Robert2016, Hannig2016}. Viewing these simulation methods within the context of CD development reveals a 
deep connection among different inferential paradigms.

\subsubsection*{Frequentist: Bootstrap estimators and CD random variables}

The bootstrap method 
simulates artificial data sets through a re-sampling process which is often referred to as {\it bootstrapping} a sample. 
The mechanism underlying the success of this simulation approach is the so-called bootstrap central limit theorem \citep{Singh1981, Freedman1981}. This theorem states general conditions under which an estimator, $\hat{\theta}$, for some parameter $\theta$ (with true value $\theta_0$) satisfies   
\begin{align}
    {\theta}^{*}_{BT} - \hat{\theta} \mid data \widesim{m} \hat{\theta} - \theta \mid \theta = \theta_0, \label{eq:boot}
\end{align}
as the sample size $n\to \infty$. In (\ref{eq:boot}), ${\theta}^*_{BT}$ represents the bootstrap estimator, that is, the estimator evaluated over a random set of bootstrapped samples. The matching statement in equation (\ref{eq:boot}) is a particular example of the approximate probability statement of equation (\ref{eq:Prob-asy}). In the bootstrapping method for inference, $\mathbb{P}_A(\cdot | data)$ is defined by the multinomial-distributed randomness of the bootstrap sampling. 
Although the multinomial-distributed randomness is typically not the same as 
the empirical randomness inherited by the statistical model, each side of equation (\ref{eq:boot}) converges to the same normal distribution by application of the central limit theorem. Therefore, (\ref{eq:boot}) can be exploited for the purpose of valid frequentist inference about $\theta$.

Consider again the setting of Example \ref{ex:norm} where $\sqrt n (\bar{Y} - \theta_0) \sim N(0,1)$ and also, by the central limit theorem, 
$\sqrt n (\bar{Y}^{*}_{BT} - \bar{Y}) \mid \bar{Y} = \bar{y} \to N(0,1)$. Here the matching expression of (\ref{eq:boot}) may be written as:  $(\bar{Y}^{*}_{BT} - \bar{Y}) \mid \bar{Y} = \bar{y} \widesim{m} (\bar{Y} - \theta)| \theta = \theta_0$, since both sides are (exactly or asymptotically) $N(0, 1/n)$.  Because
the simulated  
variability of the bootstrap estimator, conditioned upon the observed sample, matches the sampling variability of the estimator as a function of a random sample from the study population, the bootstrap method can provide valid frequentist inferential conclusions.

When the bootstrap procedure is applicable, the bootstrap estimator is closely connected to the concept of a CD-r.v. In our earlier discussion of Example \ref{ex:norm}, the CD-r.v. $\theta^*_{CD}  | \bar Y = \bar y \sim N(\bar Y, 1/n)$ 
and thus 
$({\theta^*_{CD} - \bar Y}) \big| \bar Y = \bar y \widesim{m} ({\bar Y - \theta}) \big| \theta = \theta_0$. %, where both sides follow $N(0, 1/n)$. 
Here, however, the matching is exact as in (\ref{eq:Prob-exact}) and  $\mathbb{P}_A(\cdot | data)$ is defined with respect to the Monte-Carlo randomness of simulating from a CD, rather than a resampling procedure.

In general, just as the matching in (\ref{eq:boot}) ensures the validity of bootstrap inference, a similar argument underscores the inferential validity of CD-based inference. That is 
\begin{equation}\label{eq:CDrv}
(\theta^*_{CD} - \hat \theta) \big| data \widesim{m} (\hat \theta - \theta) \big| \theta = \theta_0,
\end{equation}
where $\hat \theta$ is a point estimator (often the MLE) for $\theta$  \citep[e.g.,][]{Xie2013a}. 
Again, the matchingness of equation (\ref{eq:CDrv}) can be exact or approximate (holding as $n\to \infty$), depending on whether the CD is itself exact or asymptotic. %we are considering an exact or an asymptotic CD. 
% Comparing equations (\ref{eq:boot}) and (\ref{eq:CDrv}), 
From this, one concludes that $\theta^*_{CD}$ essentially performs in the same manner as the bootstrap estimator, $\theta_{BT}^*$. Their common trait is that the simulated artificial variability of the random estimator, $\theta^*_{CD}$ or $\theta_{BT}^{*}$, given an observed sample, matches the sampling variability of the estimator, $\hat \theta$, as a function of a random sample from the population.

In contrast to the bootstrap procedure, a CD-r.v. can be obtained through many different simulation schemes besides the multinomial re-sampling mechanism. As discussed earlier in this chapter, a CD-r.v. can be simulated from a CD that itself is  obtained through the normalized likelihood method or a $p$-value function. A CD may also be obtained from a pivotal statistic or through Bayesian or fiducial procedures, as discussed next. Therefore, although the bootstrap estimator performs similarly to a CD-r.v., a CD-r.v. is a more broadly defined concept.

\subsubsection*{Fiducial: Generalized fiducial inference and fiducial sampling} 

In modern statistical practice, R.A. Fisher's fiducial method is understood as {\it an inversion method} that solves a structural model (or algorithmic model) for some parameter $\theta$ \citep{brenner1983,Fraser1966,Fraser1968, Hannig2016}. 

Assume the data can be generated by the following algorithmic model 
\begin{equation}
    \bY = G(\theta, \bU) \label{eq:G}.
\end{equation}
Here, the form of $G(\cdot, \cdot)$ is known,
%is a general model of a known form, 
$\theta$ is the model parameter 
with true value $\theta_0$ ,
and $\bU \sim D(\cdot)$, where $D(\cdot)$ is also known but $\bU$ is an unobserved random vector of noise.  
In Example \ref{ex:norm}, the algorithmic model is $Y_i = \theta + U_i$, for $i = 1, \ldots, n$, %in the form of (\ref{eq:G}), where 
where $\bY = (Y_1, \ldots, Y_n)'$ and $\bU = (U_1, \ldots, U_n)$ follows a standard multivariate normal distribution of dimension $n$. % \sim N(0, I)$. 

In the %(now out-dated) 
classical fiducial context, model (\ref{eq:G}) is called a structural model, %known as a structural model, 
especially in the case where $\bY$ is replaced with a summary statistic 
\citep{Fraser1966, Fraser1968}. In Example \ref{ex:norm}, for instance, a structural model may be $\bar Y = \theta + \bar U$, where  $\bar Y = \sum Y_i/n$ and $\bar U = \sum U_i/n$.  Fisher suggested a  fiducial inversion process that rearranges this structural model so that 
%whereby one would solve the equation to get 
$\theta = \bar Y - \bar U$ and claimed that this equality holds for any observed $\bar Y = \bar y$ (i.e., the equation $\theta = \bar y - \bar U$ holds). %, we have $\theta = \bar Y - \bar U$. 
Since $\bar U \sim N(0,1/n)$, the {\it fiducial distribution} of $\theta$ is $N(\bar y, 1/n)$. This fiducial distribution %$N(\bar y, 1/n)$ 
is identical to the CD derived in Example \ref{ex:norm} and in this instance, it is an effective distribution estimator for inferential conclusions regarding the unknown $\theta$. However, an infamous paradox of ``hidden subjectivity'' underlies this classical fiducial argument: %, i.e., the ``hidden subjectivity'' underlying the fiducial 
the method continuously treats $\bU$ as random while conditioning upon an observed sample, $\bY=\by$ \citep{Dempster1963c}. As evident in equation (\ref{eq:G}), the data, $\bY$, and the random noise, $\bU$, are completely dependent. If one is given, the other must be as well (even if it may not be observed). Therefore, if $\bY = \by$ is observed, the correct structural equation for Example~\ref{ex:norm} is $\theta = \bar y - \bar u$, rather than $\theta = \bar y - \bar U$ where $\bar u$ is an unobserved realization of $\bar U$ corresponding to an observed realization $\bar y$ from the sampling distribution of $\bar Y$. We refer interested readers to \cite{Dawid2022} 
% (2022 -- A chapter in this BFF handbook) 
for a comprehensive review of fiducial inference in the classical fiducial inference context.

With a slight adjustment in perspective, namely by 
treating the fiducial method as an stochastic inversion algorithm that solves for a random estimator of $\theta$, modern fiducial arguments avoid the paradoxical reasoning of the classical argument and treat the fiducial distribution in a manner consistent with frequentist inference. 
 
This perspective maintains that once the data is observed, i.e. $\bY=\by$, the corresponding value of $\bU=\bu$ is also realized. Model (\ref{eq:G}) is thus replaced by   
\begin{equation}
    \by = G(\theta, \bu), \label{eq:G1}
    \end{equation}
where $\bu$ is an unobserved realization from a known distribution, $D(\cdot)$. One may simulate an artificial copy of $\bU$, say $\bu^* \sim D(\cdot)$, and use this to estimate the unobserved realization, $\bu$. A random estimator, $\theta^*$ (also known as a {\it fiducial sample}), for the parameter $\theta$ is derived by solving $\by = G(\theta^*, \bu^*)$. For repeated draws of $\bu^*$, we have multiple copies of $\theta^*$. 
The distribution of these $\theta^*$ is called a {\it fiducical distribution}. In Example \ref{ex:norm}, $\theta^* = \bar y + \bar u^*$ with $\bar u^* \sim N(0,1/n)$ and thus $\theta^* | \by \sim N(\bar y, 1/n)$; therefore, $N(\bar y, 1/n)$ is the fiducial distribution for $\theta$. As it turns out, once again %that we have the matching 
$(\theta^*  - \bar y) | \bar Y = \bar y \widesim{m} (\bar Y - \theta) | \theta = \theta_0$ and therefore the fiducial sample, $\theta^*$, plays the same role as CD-r.v., $\theta_{CD}^*$, and as the bootstrap estimator, $\bar Y^*_{BT}$.
%, a theme that will be further exploited below. 
In the fiducial context, $\mathbb{P}_A(\cdot | data)$ from equation (\ref{eq:Prob-exact}) is defined with respect to the artificial randomness of generating $\bar u^*\sim N(0,1/n)$.

Although $\theta_0$ will always solve equation (\ref{eq:G1}) for the realized pair $(\by, \bu)$, replacing $\bu$ with $\bu^*$ introduces the possibility that a solution to $\by = G(\theta, \bu^*)$ may not exist. The method of {\it generalized fiducial inference} (GFI) addresses this issue by introducing an optimization procedure to solve for
%under an $\epsilon$-approximation: 
\begin{align}
\label{eq:GFI}
    \theta_{FD,\epsilon}^* =  {\rm argmin}_{\theta \in \{\theta: \,\, || {\bf y} - G({\theta}, {\bf u}^*)||^2 \leq \epsilon \}} || {\bf y} - G({\theta}, {\bf u}^*)||^2,
\end{align} 
where $\epsilon \to 0$ at a fast rate \cite[]{Hannig2009, Hannig2016}. \citet{Hannig2009} proved a {\it Bernstein von Mises theorem} for the fiducial distribution that justifies the validity of the GFI approach, which states: under some general regularity conditions, 
\begin{align}\label{eq:fBvM}
    (\theta_{FD,\epsilon}^{*} - \hat{\theta}) \mid data \widesim{m} (\hat{\theta}-\theta) \mid \theta=\theta_0,
\end{align} 
as $n \rightarrow \infty$ and $\epsilon \to 0$ at a fast enough rate. Here, $\hat \theta$ is often the MLE. The proof of this statement involves derivations showing that both sides of equation (\ref{eq:fBvM}) converge to the normal distribution centered at zero with variance equal to the inverse of the expected Fisher information.
The matching in $(\ref{eq:fBvM})$ is approximate as defined in (\ref{eq:Prob-asy}), where $\mathbb{P}_A(\cdot | data)$
is defined with respect to $\theta^*_{FD,\epsilon}$, given observed data $\bY = \by$, % that is 
and hence with respect to the artificial variability in the simulation of $\bu^*\sim D(\cdot)$.

As with bootstrap estimators and CD-r.v.s, the validity of generalized fiducial inference hinges upon a match between the simulated variability conditioned upon the observed sample and the sampling variability under the assumed statistical model. The fiducial matching equation (\ref{eq:fBvM}) thus establishes an {\it equivalence} among the fiducial sample ($\theta^{*}_{FD}$), the bootstrap estimator ($\theta_{BT}^*$), and a CD-r.v. ($\theta^*_{CD}$) in that each estimator %. Here, the equivalence refers to that they 
plays the same role in justifying the validity of their respective inferential approaches.

\subsubsection*{Approximate Bayesian computing and Bayesian inference}

The {\it approximate Bayesian computing (ABC)} method is a Bayesian algorithm that attempts to obtain a Bayesian posterior without a direct use of Bayes' formula. It can be viewed as a Bayesian stochastic inversion algorithm to solve the same model equation (\ref{eq:G1}), where $\by$ is the observed data generated from the model (\ref{eq:G}) and $\theta$ and $\bu$ are unknown realizations from  known prior and error distributions, $\pi(\theta)$ and $D(\bu)$, respectively. The premise for establishing inferential conclusions for parameter $\theta$ in ABC is the following rejection algorithm. 

\underline{\it Approximate Bayesian Computing (ABC) Algorithm}
\begin{itemize}
 \setlength{\itemindent}{3.5em}
    \item[{\it Step 1}] Simulate $\theta^{*}\sim \pi(\cdot)$ and $\bu^{*}\sim D(\cdot)$ to compute $\by^{*} = G(\theta^*, \bu^{*}).$
    \item[{\it Step 2}] If $\by^{*}$ is equal or approximately equal to $\by_{obs}$ (i.e.,  $\by^{*} \approx \by_{obs}$), % then 
    retain $\theta^{*}$; otherwise repeat Step 1. 
\end{itemize}
Effectively, the set of retained $\theta^*$ resulting from this algorithm satisfy the equation $\by \approx G(\theta^*, \bu^*)$, so this stochastic algorithm solves equation (\ref{eq:G1}) for $\theta^*$. This inversion bears a striking resemblance to the fiducial stochastic inversion described above, except that the Bayesian inversion method assumes a prior distribution and is limited to finding those $\theta^*$ that correspond to simulations from this prior. In contrast to the $\theta^*_{FD}$ obtained from the fiducial inversion algorithm, the $\theta^*$ in the ABC algorithm contain information about both $\pi(\cdot)$ and $D(\cdot)$. 

In applications of the ABC algorithm, the approximate equality in Step 2 is difficult to achieve and is instead replaced by an approximate equality of some summary statistics while permitting a small degree of mismatch, $\epsilon > 0$. Thus, in practice, Step 2 of the ABC algorithm is replaced with Step 2' below, where $d(\cdot, \cdot)$ is a distance metric (often Euclidean distance), $t(\cdot)$ is a summary statistic:
%and $\epsilon > 0$ is the tolerance for mismatch: 
\begin{itemize}
\setlength{\itemindent}{3.5em}
\item[{\it Step 2'}] If $d(t(\by^*), t(\by_{obs})) < \epsilon$, retain $\theta^{*}$; otherwise repeat Step 1. 
\end{itemize}

Given the observed data, the underlying distribution of the output $\theta^*$ from the ABC algorithm is called the {\it ABC posterior}, which we denote by $f_{a}(\theta \mid \by)$. If the summary statistic $t(\cdot)$ is sufficient, then the ABC-posterior (approximately) matches the target posterior distribution, $f(\theta \mid \by)$, derived from Bayes formula, when $\epsilon \to 0$; cf, e.g., \citet{Frazier2018}; \citet{Thornton2019}. In such circumstances, the ABC method is a computational approach %that can be used to 
to obtain the posterior distribution without the use of Bayes' formula or a direct evaluation of the likelihood function. Note that, when given the observed data $\by$, the randomness in $\theta^*$ is induced by artificial Monte-Carlo simulations of $\theta^* \sim \pi(\cdot)$ and $\bu^* \sim D(\cdot)$. 

The {\it Bernstein von Mises theorem} for the posterior distribution is a well-known  
bridge relating Bayesian and frequentist inference for large  sample sizes, e.g., \citet{Vandervaart1998}. Denote by
$\theta_0$ the realized value of $\theta$ from the prior %is generated, 
and let $\theta_{BY}^{*}$ be an outcome from an ABC algorithm, or, more generally,
a Monte-Carlo sample draw from a posterior distribution, given 
the observed data $\by$. We can restate the Bernstein von Mises theorem in this context as follows:

{\it Under some general regularity conditions and as $n \rightarrow \infty$,
\begin{equation} \label{eq:bBvM}
(\theta_{BY}^{*} - \hat{\theta}) \mid data \widesim{m} (\hat{\theta}-\theta) \mid \theta=\theta_0,    
\end{equation}
where the above follow a normal distribution centered at zero with variance equal to the inverse of the expected Fisher information, and  $\hat{\theta}$ is (often) the MLE.}

\noindent The underlying probability measure, $\mathbb{P}_A(\cdot | data)$, on the left hand side of equation (\ref{eq:bBvM}) is defined with respect to the %(approximate) 
posterior probability. If $\theta_{BY}^{*}$ is instead a Monte-Carlo draw from the posterior distribution exactly, then $\mathbb{P}_A(\cdot | data)$ is defined with respect to %is equivalently with respect to 
the artificial randomness of the Monte-Carlo simulation only.  
In the ABC algorithm presented above, %example of the ABC method mentioned in this paper, 
the randomness of $\mathbb{P}_A(\cdot | data)$ is derived from %then for 
the randomness of
the Monte-Carlo sampling of $\theta^*$ and $\bu^*$ in Step 1.  

ABC methods are valued %often billed 
as a Bayesian approach that does not require the %without a 
direct specification of the likelihood function (or direct use of Bayes' formula). %, and they 
As such, these methods are most useful in applications where one has little to no knowledge about the sufficiency (or near sufficiency) of the summary statistic. If the summary statistic, $t(\cdot)$, %used in an ABC algorithm 
is not a sufficient statistic, then the ABC posterior distribution can be quite different from the targeted posterior distribution as demonstrated in the Cauchy example discussed next. Thus, considerations for the inferential validity of conclusions based on $f_{a}(\theta \mid \by)$ for insufficient
% non-sufficient \ST{(insufficient?)} 
$t(\cdot)$ are critical to consider for any practical use of the ABC algorithm. 
To this end, \citet{Thornton2019,Thornton:arxiv} derive some regularity conditions under which the matching equation~(\ref{eq:bBvM}) still holds either exactly (as in (\ref{eq:Prob-exact})) or asymptotically (as in (\ref{eq:Prob-asy})), and thus an ABC posterior, $f_{a}(\theta | \by)$, is %often 
arguably a valid CD, regardless of whether or not $t(\by)$ is sufficient. Example \ref{ex:cauchy} explores such a case.

\begin{example}\label{ex:cauchy}
Suppose we observe $\by=(y_1, y_2, \dots, y_n)$, an independent sample from a $Cauchy(\theta,1)$ distribution. We would like to use the ABC algorithm to obtain ABC posteriors $f_{a}(\theta \mid \by)$ using a flat prior. We consider two popular summary statistics, the sample mean, $t_1(\by) = \bar y$, and the sample median, $t_2(\by) = y_{(m)}$.
%, (here $m = (n+1)/2$ if $n$ is odd or $n/2$ if $n$ is even). 
Since, the distribution of the Cauchy sample mean follows a $Cauchy(\theta,1)$ distribution, one can show that $f_{a}(\theta \mid \by)$ is $Cauchy(\bar y,1)$, as $\epsilon \to 0$ (for any given finite $n$) \citep[e.g.,][]{Thornton2019}. Thus for the ABC method based on the summary statistic $\bar y$, the matching equation
$(\theta_{BY}^* - \bar Y) | \bY = \by \widesim{m} (\bar Y - \theta) | \theta= \theta_0$ holds for finite $n$ and as $\epsilon \to 0$.  Similarly, 
since the distribution of the Cauchy sample median follows a $N(\theta, \pi^2/4n)$ distribution as $n \to \infty$, one can show that $f_{a}(\theta \mid \by)$ is asymptotically $N(y_{(m)}, \pi^2/4n)$ as well. Once again, for the ABC method based on the summary statistic $y_{(m)}$, the matching equation
$\theta_{BY}^* - Y_{(m)} | \bY = \by \widesim{m} (Y_{(m)} - \theta) | \theta= \theta_0$, holds as $n \to \infty$ and $\epsilon \to 0$. (The rate at which $\epsilon \to \infty$ is dependent on the sample size and must be faster than $O(n^{-1/2})$ \citep{li2018}.) 
In both cases, $\mathbb{P}_A(\cdot | data)$ on the left hand sides of the matching equations represents the simulation randomness of the Monte-Carlo draws in the ABC algorithm. A particularly noteworthy feature of this example, illustrated in Figure \ref{fig:cauchy}, is that neither summary statistic, $t_1$ nor $t_2$, is sufficient and both corresponding ABC posteriors fail to converge to the actual posterior of the target parameter $\theta$. 
However, because both summary statistic choices yield a distribution such that equation (\ref{eq:bBvM}) holds, \citet{Thornton2019} uses the CD definition to argue that the ABC posteriors 
are both CDs. Therefore, although the ABC algorithm does not lead us to the traditional Bayesian inference for either $t_1$ or $t_2$, the output of the ABC algorithm is still useful for valid (frequentist) inference. 

Despite the dramatic difference in efficiency of these two example CDs, the inferential validity of statistical conclusions from either method remains intact according to the definition of a CD. This conclusion is demonstrated for a sample of size $n=40$ in Table \ref{tab:cauchy}. 
See \citet{Thornton2019,Thornton:arxiv} for more discussion on the role of equation (\ref{eq:bBvM}) in establishing a CD.

\begin{figure}[h] 
\begin{center} 
	\caption{{\it Given a sample of size $n=40$ from a $Cauchy(\theta,1)$ distribution, these are the densities for both ABC posterior distributions (in black) and for the target posterior distribution (in gray). Here we set $\theta_0 =10$ and select a flat prior, $\pi(\theta) \propto 1$. Note that the much tighter density curve for the ABC posterior based on the sample median indicates that this method is far more efficient than the ABC posterior based on the sample mean. 
	}}\label{fig:cauchy}
	{\includegraphics[width=.8\linewidth, height=4in]{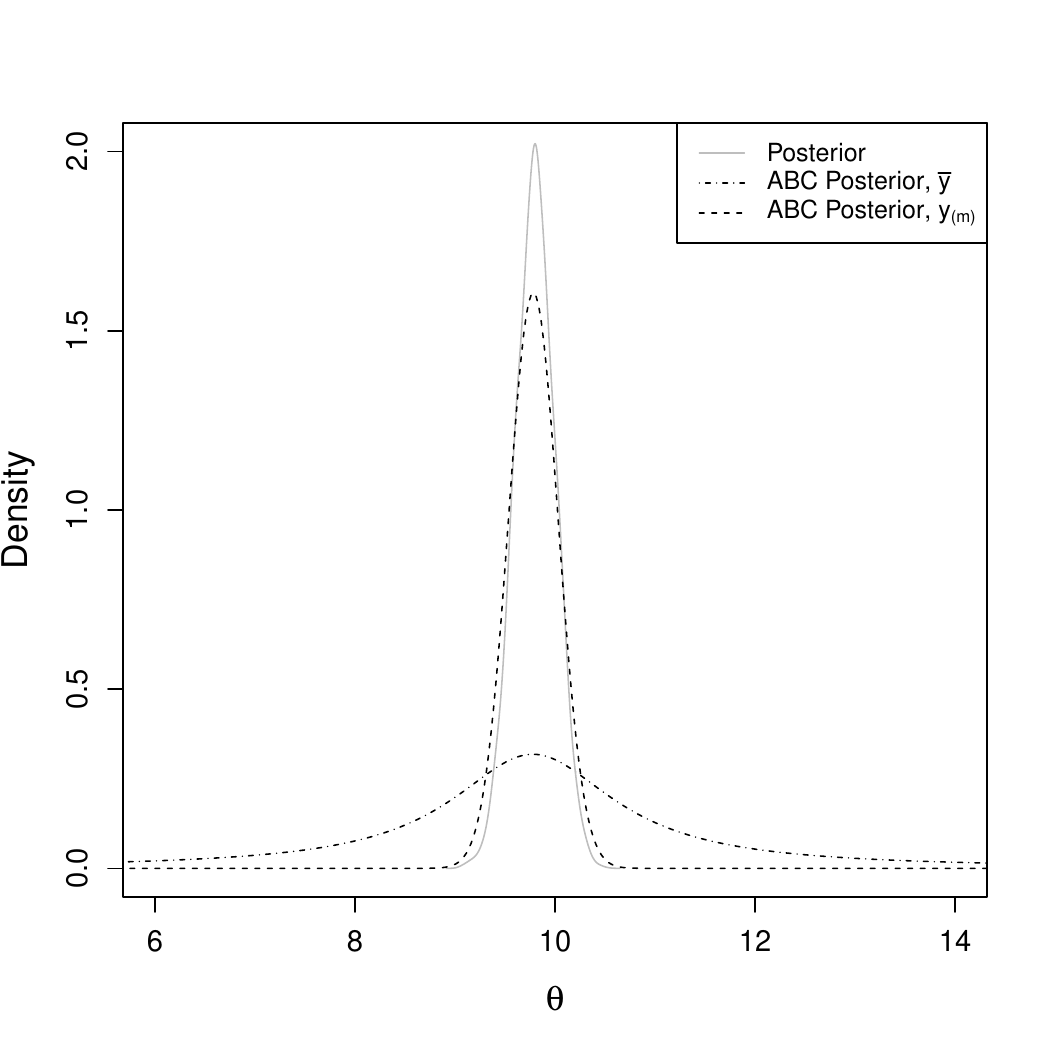}} 
\end{center} 
\end{figure}	

\begin{table}[h]
\caption{{\it Coverage results over $800$ random samples for the two ABC posterior distributions and the target posterior distribution in the same settings as Figure \ref{fig:cauchy}. The width of the confidence intervals from the ABC posterior based on the sample median is much smaller indicating this method is (far) more efficient than the other method based on the sample mean. 
} }
\label{tab:cauchy}
\begin{center}
\begin{tabular}{rccr} \hline\hline
 & & \multicolumn{2}{c}{$95\%$ confidence or credible intervals} \\
 \cline{3-4}
    {Method} &   & Coverage         &     Width       \\ \hline 
ABC posterior with $t({\bf y}) =\bar{y}$ &   &    0.934         &     25.412      \\  
ABC posterior with $t({\bf y}) = y_{(m)}$  & &    0.944         &     0.974       \\ 
Bayesian posterior   &             &    0.964         &     0.897 \\
\hline
\end{tabular}
\end{center}
\end{table}
\end{example}

As indicated in the above Cauchy sample mean case, (\ref{eq:bBvM}) can also hold without relying on the (Bayesian) Bernstein von Mises theorem. 
The structure of (\ref{eq:bBvM}) is analogous to (\ref{eq:boot}), (\ref{eq:CDrv}) and (\ref{eq:fBvM}); the only 
distinction is that Bayesian inference imposes an additional model assumption (the prior) and this information 
factors into the inferential conclusions along with the model uncertainty of $\bU$. The common conclusion from all four of these analogous statements is that the random estimators, $\hat \theta_{BT}^{*}$, $\theta_{CD}^{*}$, $\theta_{FD}^{*}$ and $\theta_{BY}^*$, behave according to the same standard of frequentist inference. These random estimators are useful precisely because of this common trait: the artificial Monte-Carlo randomness incorporated into each of these estimators matches the sampling variability of $\hat \theta$.  
Thus we have the final piece of a puzzle that connects  
Bayesian, frequentist and fiducial inference through the construction of random estimators ($\theta_{BY}^*$, $\theta_{FD}^*$, and $\theta_{BT}^*$) that can all be described as  CD-random variables, $\theta^*_{CD}$.

Finally, to conclude this subsection, we remark that the matching equations (\ref{eq:boot}), (\ref{eq:CDrv}), (\ref{eq:fBvM}) and (\ref{eq:bBvM}) rely on the first-order large sample theory. 
%bootstrap CLT and fiducial and Bayesian versions of a Bernstein von Mises theorem. 
The large sample Bernstein von Mises theorem for the posterior distribution mitigates the impact of the prior on inference in favor of growing data to make this alignment possible.  
These matching equations can be extended to higher-order results (e.g., higher-order results for the boostrap method in \citet{Hall1992}, for likelihood-based methods in \citet{Reid2010}, and for objective Bayesian methods in \citet{Berger:Bernardo:Sun2015}). 
Nevertheless, these matching equations can also be exact as in (\ref{eq:Prob-exact}), leading to exact inference results that are valid under a finite sample size. Beside the non-normal matching example of the Cauchy sample using $t_1 = \bar{y}$ in Example \ref{ex:cauchy}, there are some other recent works that utilize the exact matching (\ref{eq:Prob-exact}) to make finite-sample inferential statements where neither the central limit theorem nor a high order asymptotic development is required; e.g., \citet{LiuS2018}.

%------------------------------------

\subsection{Duality of parameters in BFF 
inference}\label{sec:params}
In statistics, we have traditionally distinguished inferential perspectives by asserting that the model parameter is a fixed, unknown constant in frequentist inference but is random in Bayesian inference. From this perspective, these inferential paradigms seem to conflict. Can we reconcile these different understandings of a parameter to create a meaningful bridge between the information gained from either  perspective? Our answer to this question is an emphatic ``yes'' as we argue in the remainder of this subsection that there are two reconcilable versions of the parameter, random and fixed, in each of the Bayesian, frequentist, and fiducial inferential paradigms. We will use our understanding of the common trait of random estimators and the matching equations from the previous subsection to support this argument.

Let's start our discussion by considering {\it Bayes’ Billiard Table Experiment}, from whence the field of statistics was developed:   
\begin{itemize}
\item[]  {\it Assume that billiard balls are rolled on a line of length one, with a uniform probability. Ball $W$ is rolled and stops at $\theta_0 \in (0,1)$. Then a different ball $O$ is rolled $n$ times under the same assumptions. Let $y$ be the number of times that ball $O$ stops to the left of ball $W$. 
Bayes proposed the motivating question: ``Given $y$, what inference can we make concerning $\theta_0$?''\cite[]{Bayes1764}}
\end{itemize}
In modern statistical terms, Bayes’ Billiard Table experiment assumes
a $U(0,1)$ prior distribution and collects a binomial sample of data. In this context, $\theta_0$ (where ball $W$ landed) is a realization from the prior distribution, $\theta \sim U(0, 1)$, and the sampling scheme follows a binomial model $Y|\theta~=~\theta_0~\sim~Bin(n, \theta_0)$. For the sake of discussion and without loss of generality, suppose ball $W$ lands at the location $\theta_0= 0.3896$ and suppose out of the $n =14$ times ball $O$ was rolled, we observe $y = 5$ instances where $O$ landed to the left of ball $W$. This $y = 5$ is a realization from $Y| \theta = 0.3896 \sim Bin(n = 14, \theta = 0.3896)$.  We emphasize that the target quantity we are interested %in estimating
is the fixed, unknown value $\theta_0 =0.3896$, the location where ball $W$ landed, producing %resulted in 
the observation $y = 5$. By Bayes' formula,  the posterior distribution is $\theta | y = 5 \sim Beta(6, 10)$. In this experiment, the target of interest is not a random quantity that follows either a $U(0,1)$ or a $Beta(6, 10$) distribution, rather it is a fixed value for parameter $\theta$ that resulted in the observation of $y=5$.

Bayes’ Billiard Table experiment clearly displays two versions of the parameter at play, which we call the  
{\it duality of a parameter}.  
The scientific, application-oriented point of view, is interested in the fixed version of the parameter that generated the observed data, i.e. $\theta_0$. There is uncertainty however in exactly how to assess $\theta_0$; to address this uncertainty, a Bayesian inferential approach elects to work with a random version of $\theta$ through the prior and posterior distributions.
As stated in \citet{Berger:Bernardo:Sun2015}, in a Bayesian perspective “parameters must have a distribution describing the available information about their values ... this is not a description of their variability (since they are fixed unknown quantities), but a description of the uncertainty about their true values.”

The duality of a parameter is actually present in other inferential paradigms as well. The model parameter has two versions: (a) a fixed, unknown true (or realized) value which generated the observed data, and (b) a random version that is used to describe the uncertainty (but not the variability) %about 
in estimating the fixed, unknown parameter value. The duality of parameters was first recognized in the fiducial argument, although the philosophical underpinnings of the fiducial interpretation have probably generated more confusion and controversy than clarity (see, e.g. \citealp{rao1973linear}). From a frequentist perspective,
modern developments regarding CDs, as summarized in the previous subsection, support lucid arguments for considering a random version of the model parameter in tandem with a fixed, data-generating version. In particular, a CD-r.v. $\theta^*_{CD}$ can be viewed a random version of the parameter. As demonstrated in the previous subsection, $\theta^*_{CD}$
plays the same role as the Bayesian random parameter, $\theta_{BY}^*$, and a fiducial sample, $\theta^*_{FD}$, and 
the variability in $\theta^*_{CD}$ matches the empirical uncertainty inherited from the assumed population model.

%\ST{Table X summarises how these two versions of a parameter apply across each inferential paradigm. In Table X, the fixed version of the parameter is the target of interest from which the sample of data is assumed to have been generated.} 
Altogether, these two versions of a parameter apply across each inferential paradigm. The fixed version of the parameter is the target of interest from which the sample of data is assumed to have been generated. It has the same standard interpretation in frequentist and fiducial inference and it refers to the realized parameter value, $\theta_0$, from the prior under the Bayesian framework.  The random version is used to describe the uncertainty in our inference about the fixed version. In Bayesian inference this has a natural interpretation, but we argue that we can also adopt the same interpretation for the corresponding CD-r.v. (or bootstrap estimator in special cases) and fiducial sample in the frequentist and fiducial frameworks.

%\begin{table}[ht]
%\centering \caption{
%Parameter duality among different inferential paradigms.}
%\resizebox{.95\linewidth}{!}{\begin{tabular}{rcc}
%\hline\hline
%& \textbf{Random Version} (describing uncertainty) & \textbf{Fixed Version} (describing target) \\ \hline
%\textbf{CD}   & CD-r.v. ($\theta^*_{CD}$) & True parameter value ($\theta_0$) \\ \textbf{Bootstrap}   &  Bootstrap estimator ($\theta^*_{BT}$) & True parameter value ($\theta_0$) \\ \textbf{Fiducial}  & Fiducial sample ($\theta^*_{FD}$)  & True parameter value ($\theta_0$) \\  \textbf{Bayesian}   & Random parameter (prior/posterior sample) ($\theta^*_{BY}$) & Realized parameter value ($\theta_0$)                    \\ \hline
%\end{tabular}}
%\label{tab:common}
%\end{table}

Whereas much previous research has highlighted differences among Bayesian, frequentist and fiducial inference, 
this unifying perspective % inspired by CD 
can deepen our understanding of the foundational principles of statistical inference and provides a philosophical framework whereby potentially any inferential method can benefit from each of the three dominant paradigms. For instance, recognizing and establishing the random version of parameters in frequentist and fiducial inference suggests that the powerful statistical computational tools that have been so successful in many Bayesian applications can also be applied to frequentist and fiducial inference. Alternatively, introducing the interpretation of a fixed target parameter in the context of Bayesian inference can help directly and unambiguously connect the interpretation of parameters to the physical meanings of parameters in physics and other applied scientific domains. Furthermore, the CD-based connection among bootstrap sampling and approximate Bayesian computing reinforce the usefulness of developing new sampling methods to address more difficult inference problems that lie beyond the reach of likelihood-based inference and extensions of the central limit theorem % (and its extensions),
\citep[e.g.,][]{LiuS2018, 
%Cui2019, 
Luo2020}.

\section{Further Discussion and Concluding Remarks}\label{sec:discussion}

A CD is a sample-dependent distribution function on the parameter space that can represent confidence intervals (or regions) of all levels for some parameter of interest.  The CD approach summarizes the information available from the observed data in a distribution form. %, when possible. 
An emerging theme from the development of CD estimators %the development %of CD estimators 
is that any approach, regardless of whether it is Bayesian, fiducial, or frequentist, can potentially be unified under the concept of CDs as long as the method can be used to build confidence intervals (regions) of all levels, exactly or asymptotically. This union % unification 
provides a platform to compare and also combine a variety of distribution estimators %``distribution estimates'' 
that may have been derived through various procedural methods. 
Furthermore, we have also
demonstrated that each of the frequentist, fiducial, and Bayesian inferential frameworks share two important features: an ability to describe parameter uncertainty with a random version and a view that there is some fixed, unknown value as the target of interest. The random version of the parameter is associated with distribution estimators, namely, a posterior distribution, fiducial distribution or CD, across all inferential paradigms. When the model is given, the common theory critical to the success of each of these inferential frameworks aligns the variability of the random estimator (conditional upon the observed sample) with the
empirical uncertainty about the fixed version of the parameter. These key similarities hint at the broad range of inference problems that may be solvable by simulating Monte-Carlo samples that mimic the observed data.

In addition to supporting inferential unity among different statistical paradigms, this CD-based perspective promises many new methodological developments that can provide novel inferential~solutions in cases where traditional solutions were previously impossible. This potential for new developments includes prediction \citep[e.g.,][]{Shen2018,Vovk2019,Xie2021, Tian2021}, testing methods and simulation schemes \citep[e.g.,][]{Liu2020, Luo2020}, and combining information from various inferential sources \citep[e.g.,][]{Xie2011,Hannig2012, Clagget2014,Liu2015a,Shen2020, Cunen2021, Cai2020}.

We remark that the two versions of randomness explored in this chapter do not have to mirror the classical discussions of aleatory and epistemic probabilities.
Rather, one may connect the uncertainty in the probability statement $\mathbb{P}(\cdot)$ on the right hand side of (\ref{eq:Prob-exact}) and (\ref{eq:Prob-asy}), with the uncertainty of randomly sampling % data 
from an assumed population. The latter represents empirical variability, and apparently can also be explained as an aleatory probability. % However, t
On the other hand, the Monte-Carlo induced randomness, described by $\mathbb{P}_A(\cdot | data)$, does not necessarily have a straightforward interpretation as an epistemic probability.
Although we may argue that the Monte-Carlo randomness is intended to describe the uncertainty in knowledge, it is derived from a concrete Monte-Carlo procedure that mimics the observed sampling mechanism and does not have an obvious interpretation as ``a measure of the `degree of belief' of the individual assessing the uncertainty of a particular situation'' \citep[chap. 7]{ramsey1931}.
The so-called ``personal belief'' is not under consideration in this paper,  since, when a prior is assumed, the desired matching that we promote either requires a matching prior or that the information in the prior can be ``washed out'' by a large sample size through the Bernstein von Mises theorem or its generalizations. 
When the prior is determined from historical data or knowledge,  %there is an
one may still attempt to model the uncertainty both in the prior and assumed model through CDs, e.g., \citet{xie2013incorporating}. In this case, the epistemic probability discussed in a typical Bayesian analysis is replaced by an empirical (aleatory) probability inherited from the prior model (historical experiments) and the observed sample. Finally, there are different views on how to interpret the probability 
underscored by the CD. For instance, \citet[chap. 15]{Schweder2016} take the position that the probability described by a CD is epistemic, which differs from our perspective. 
Our perspective is more aligned with that offered in \citet[\textsection 2.2]{Reid2015}: ``We may avoid the need for a different version of probability by appeal to a notion of calibration, measured by the behaviour of a procedure under hypothetical repetition.''
More discussions on this aspect and other principles of statistical inference can be found in \citet{Reid2015}. 

Although this chapter emphasizes the unity among the Bayesian, frequentist and fiducial inferences, these procedures are distinct in several respects. A Bayesian method includes an additional model assumption through the prior distribution. Depending on the situation, this additional assumption may result in inferential conclusions that do not match analogous frequentist conclusions, especially in the case of a finite sample size or higher order asymptotic developments, e.g., \citet{Fraser2011, Reid2015}. Furthermore, the conclusions reached through a fiducial inversion procedure or through a Bayesian approach do not automatically (even if they do typically) lead to the correct frequentist inference with respect to a CD. This issue is similar to the fact that an MLE is usually but not automatically a consistent estimator and also that a consistent estimator does not need to be an MLE. (See Table \ref{tab:behaviorist}.) Nevertheless, the differences among Bayesian, frequentist and fiducial procedures do not negate underlying congruencies among these methods namely, a practical duality in the interpretation of model parameters and an inferential connection through confidence distributions.

\bibliographystyle{chicago}
\bibliography{CD-BFF-Handbook}
\printindex
\end{document}